# A study of MIR photoluminescence from Pr$^{3+}$ doped chalcogenide fibers pumped at near-infrared wavelengths

S. Sujecki[*,1,2], L. Sojka[1], E. Beres-Pawlik[1], R. Piramidowicz[3], H. Sakr[2], Z. Tang[2], E. Barney[2], D. Furniss[2], T.M. Benson[2], A.B. Seddon[2]

[1]Department of Telecommunications and Teleinformatics, Faculty of Electronics, Wroclaw University of Science and Technology, Wyb. Wyspianskiego 27, 50-370 Wroclaw, Poland

[2]George Green Institute for Electromagnetics Research, the University of Nottingham, University Park, NG7-2RD, Nottingham, UK

[3]Institute of Microelectronics and Optoelectronics, Warsaw University of Technology Nowowiejska 15/19, 00-665 Warsaw, Poland

**ABSTRACT**

We perform a numerical analysis of mid-infrared photoluminescence emitted by praseodymium (III) doped chalcogenide selenide glass pumped at near-infrared wavelengths. The results obtained show that an effective inversion of level populations can be achieved using both 1480 nm and 1595 nm laser diodes. The rate of the spontaneous emission achieved when pumping at 1480 nm and 1595 nm is comparable to this achieved using the standard pumping wavelength of 2040 nm.

## 1. INTRODUCTION

Mid infrared (MIR) light chalcogenide fiber based photoluminescence sources find increasingly many applications in the sensor technology. Dysprosium (III) and praseodymium (III) doped sulfide chalcogenide glass based sources were used for gas [1] and water pollutant [2] sensing. In this contribution we explore the luminescence properties of selenide-chalcogenide glass fibers doped with praseodymium (III) ions for potential application as MIR spontaneous emission sources for MIR light based sensors. For pumping purposes we consider an application of near infrared laser diodes, which are robust, reliable and low cost. A large research effort was invested into the development of the lanthanide ion doped chalcogenide-selenide fiber lasers. Numerical results predict that an efficient operation of such devices is possible [3]. Also, strong mid-infrared photoluminescence was experimentally obtained from praseodymium (III) doped both large and small core chalcogenide-selenide glass fibers [4-6]. This makes praseodymium (III) doped chalcogenide-selenide glass fiber a good candidate for the realization of fiber based mid-infrared spontaneous emission sources.

## 2. MODELLING AND RESULTS

For the purpose of this study we fabricated several samples of selenide-chalcogenide glass doped with praseodymium ions and performed absorption and photoluminescence spectrum measurements. From the experimental results using the standard procedures we extracted the relevant modelling parameters [7].

The population inversion in Pr$^{3+}$ ion doped selenide-chalcogenide glass can be studied using a rate equations model [8]. Here we considered 4-level system corresponding to the pump wavelength of approximately 1.5 μm.

Figure 1 shows the energy level diagram relevant for praseodymium ions when a near-infrared pump operating around 1.5 μm is used to populate the higher energy levels. Due to a fairly small energy gap when compared with the maximum phonon energy of the chalcogenide-selenide glass host the level 4 is depopulated mainly non-radiatively.

*slawomir.sujecki@pwr.edu.pl

The rate equations for the 4-level system shown in Fig.1 consist of four algebraic equations:

$$\begin{bmatrix} 1 & 1 & 1 & 1 \\ 0 & a_{22} & a_{23} & a_{24} \\ 0 & 0 & a_{33} & a_{34} \\ a_{41} & 0 & 0 & a_{44} \end{bmatrix} * \begin{bmatrix} N_1 \\ N_2 \\ N_3 \\ N_4 \end{bmatrix} = \begin{bmatrix} N \\ 0 \\ 0 \\ 0 \end{bmatrix} \quad (1)$$

where the matrix elements $a_{xx}$ are:

$$a_{22} = -\left(\frac{1}{\tau_2} + \frac{1}{\tau_{21}^{mp}}\right); a_{23} = \frac{\beta_{32}}{\tau_3} + \frac{1}{\tau_{32}^{mp}}; a_{24} = \frac{\beta_{42}}{\tau_4}; a_{33} = -\left(\frac{1}{\tau_3} + \frac{1}{\tau_{32}^{mp}}\right); a_{34} = \frac{\beta_{43}}{\tau_4} + \frac{1}{\tau_{43}^{mp}};$$

$$a_{41} = \sigma_{14}^a \phi_p; a_{44} = -\left(\sigma_{41}^e \phi_p + \frac{1}{\tau_4} + \frac{1}{\tau_{43}^{mp}}\right);$$

$\quad (2)$

In (2) $\sigma^a_{14}$ and $\sigma^e_{41}$ are the relevant values of absorption and emission cross sections, respectively. $\beta_{xx}$ give the relevant values of the branching ratios. $\tau_2$, $\tau_3$ and $\tau_4$ are radiative lifetimes of levels 2, 3 and 4, respectively while $\tau_x^{mp}$ give the relevant life times for phonon assisted transitions [4], $\tau_{xr}$ and transitions, respectively.

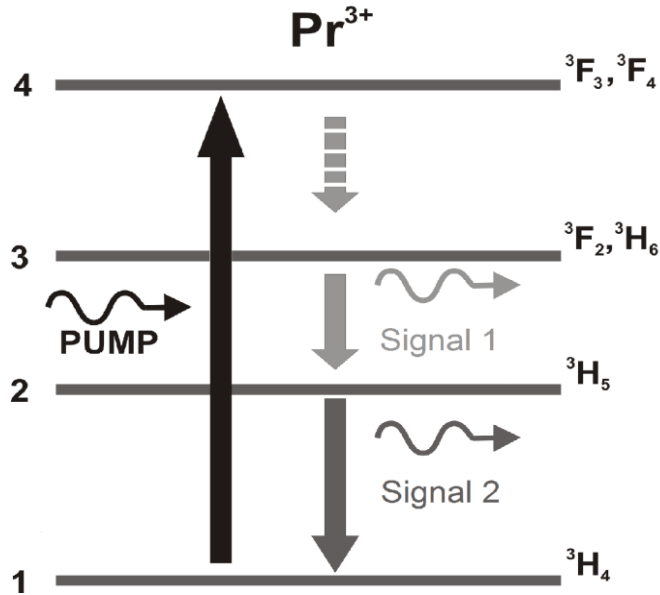

Figure 1. Energy level diagram

Figure 2 shows the dependence of the emission and absorption cross sections on the wavelength. There are three maxima of the absorption cross section within the presented wavelengths range: 1480 nm, 1595 nm and 2040 nm. From the practical point of view the particularly useful is the maximum at 1480 nm due to availability of low cost high power laser diodes for this wavelength. Therefore Fig. 3 shows the dependence of the relevant level populations on the pump intensity for the considered 4 level system when the pump wavelength was set at the 1480 nm. The assumed luminescence life time of level 3 is 4 ms while for the level 2 is 12 ms. The multiphonon transition lifetime for the levels 4, 3 and 2 is 5 μs, 9 ms and 9 ms, respectively. Fig.3 shows that a significant inversion of the level 3 and 2 populations with respect to the ground state is achieved for pump intensity exceeding 1 MW/m². It can be also observed that due to fairly high multiphonon transition rate level 4 is practically empty. For comparison in Figs. 4 and 5 we showed the dependence of the level populations on the pump intensity when pumping at 1595 nm and 2040 nm, respectively. These results show that a similar degree of level population inversion is achieved at the corresponding value of the pump intensity when compared with 1480 nm pumping. Thus the only advantage of pumping using the longer wavelengths is a smaller quantum defect.

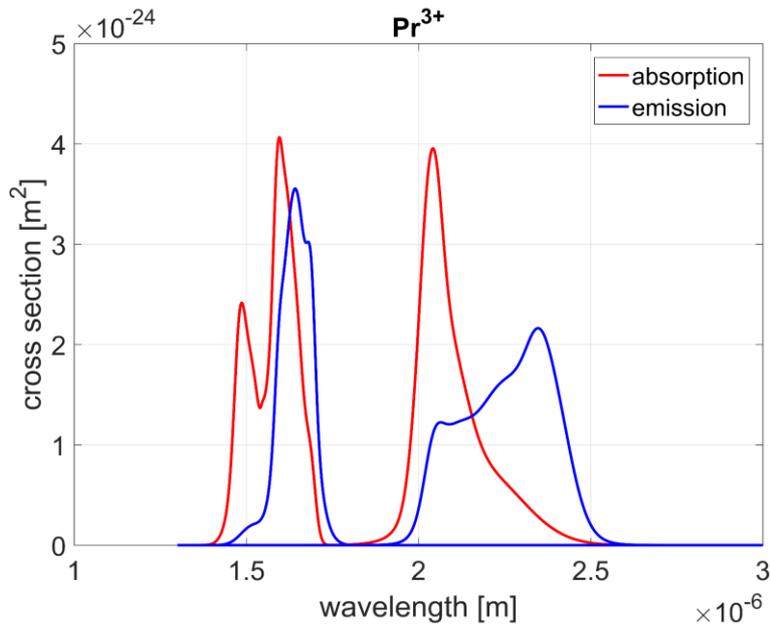

Figure 2. Measured emission and absorption cross section for $Pr^{3+}$ doped chalcogenide-selenide glass

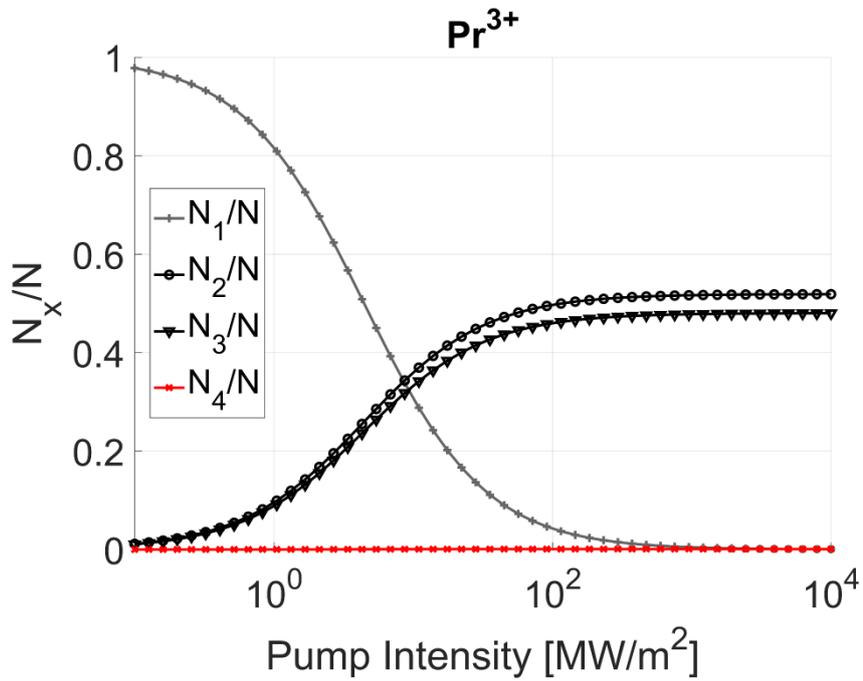

Figure 3. Dependence of the level populations on the pump intensity for $Pr^{3+}$ doped chalcogenide-selenide fiber pumped at 1.48 μm.

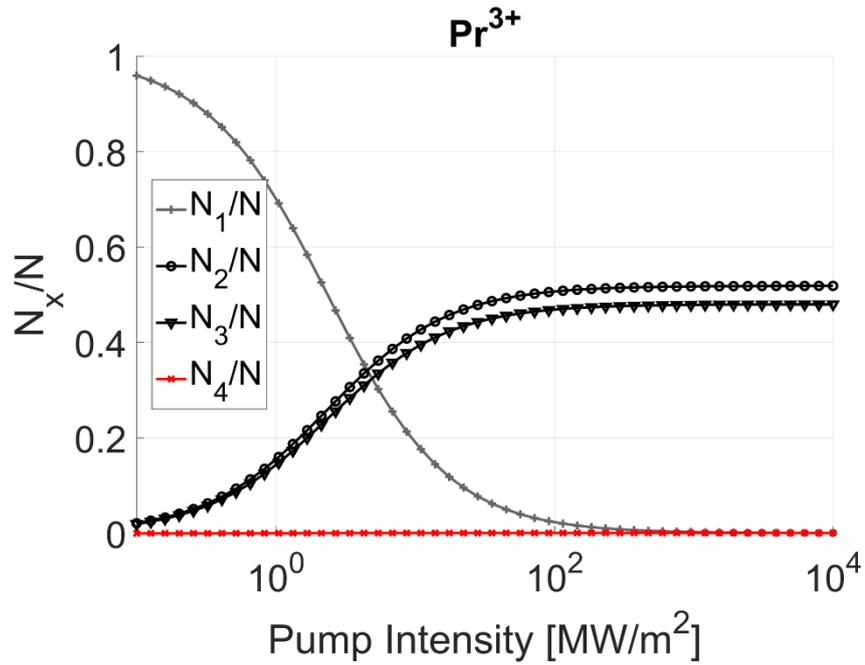

Figure 4. Dependence of the level populations on the pump intensity for $Pr^{3+}$ doped chalcogenide-selenide fiber pumped at 1.595 μm.

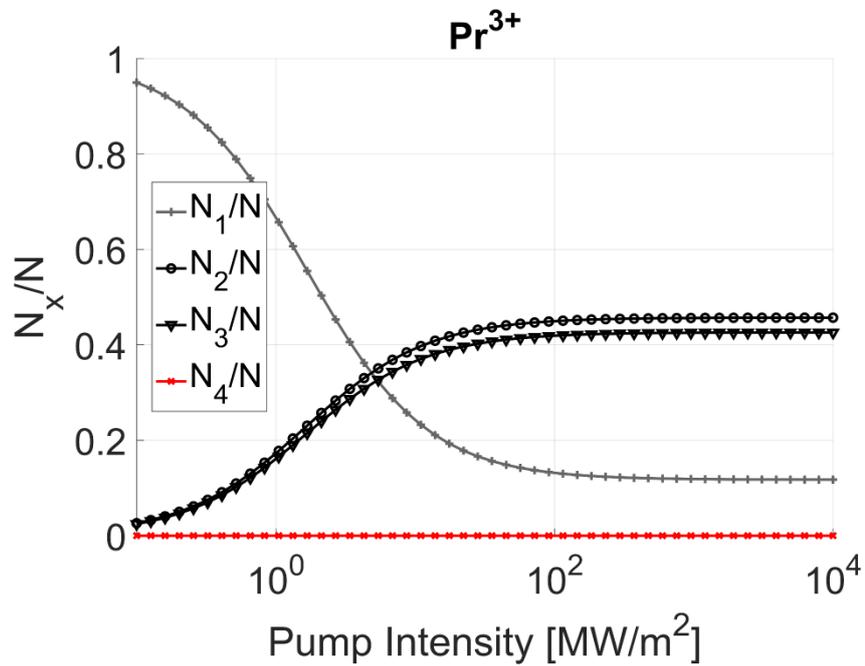

Figure 5. Dependence of the level populations on the pump intensity for $Pr^{3+}$ doped chalcogenide-selenide fiber pumped at 2.04 μm.

In Figs. 6, 7 and 8 we compare the dependence of the spontaneous emission (photoluminescence generation) rate on the pump intensity for 1480 nm, 1595 nm and 2040 nm pumping wavelengths, respectively. Again similar values of the spontaneous emission rate at the corresponding value of the pump intensity are obtained in all three cases. These results further support the claim that 1480 nm pumping can be effectively used instead of the 1595 nm or 2040 nm pumping.

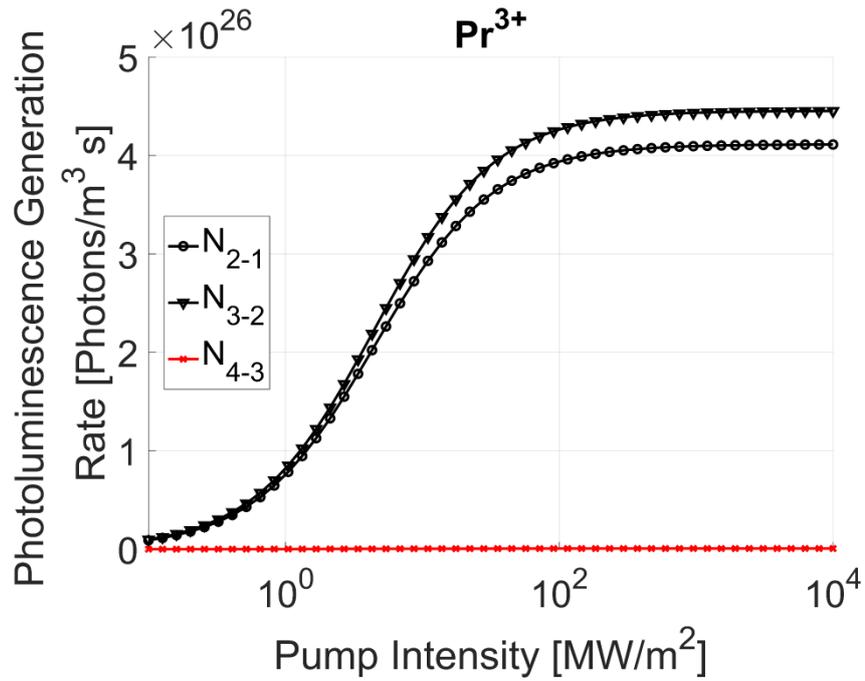

Figure 6. Dependence of the photoluminescence generation rate on the pump intensity for $Pr^{3+}$ doped chalcogenide-selenide fiber pumped at 1.48 μm.

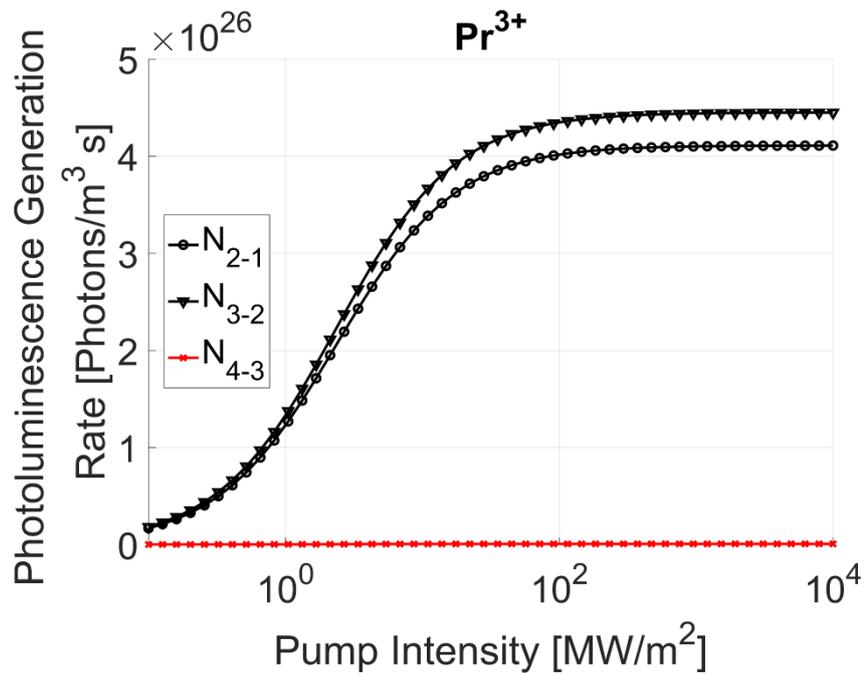

Figure 7. Dependence of the photoluminescence generation rate on the pump intensity for $Pr^{3+}$ doped chalcogenide-selenide fiber pumped at 1.595 μm.

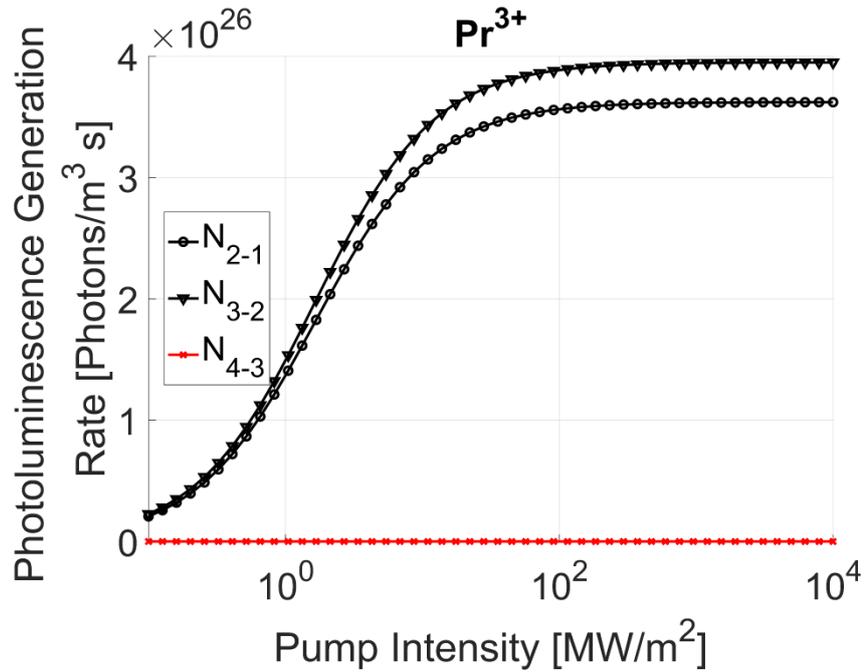

Figure 8. Dependence of the photoluminescence generation rate on the pump intensity for $Pr^{3+}$ doped chalcogenide-selenide fiber pumped at 2.04 μm.

## 3. CONCLUSIONS

We investigated the effect of using near infrared wavelengths for pumping chalcogenide-selenide. The study shows that 1480 nm pumping is nearly as effective as the 1595 nm and 2040 nm pumping. The only significant disadvantage of 1480 nm pumping when compared with longer wavelengths is a larger value of the quantum defect.

## ACKNOWLEDGEMENTS

This project has received funding from the European Union's Horizon 2020 research and innovation programme under the Marie Skłodowska-Curie grant agreement No. 665778 (National Science Centre, Poland, Polonez Fellowship 2016/21/P/ST7/03666).